\documentclass[12pt,preprint]{aastex}

\shorttitle{Einstein ring}
\shortauthors{Ghosh \& Narasimha}

\begin{document}

\title{A QSO plus host system lensed into a 6" Einstein ring  by a low redshift galaxy}


\author{Kajal K Ghosh\altaffilmark{1}}
\affil{USRA/NSSTC/MSFC/NASA, 320 Sparkman Drive, Huntsville, AL 35805, USA}
\email{kajal.k.ghosh@nasa.gov}

\and

\author{D. Narasimha\altaffilmark{2}}
\affil{Tata Institute of Fundamental Research, Mumbai 400 005, India}
\email{dna@tifr.res.in}


\begin{abstract}
We report the serendipitous discovery of an ``Einstein Ring" in the optical
band from the Sloan Digital Sky Survey (SDSS) data and associated four
images of a background source.
The lens galaxy appears to be a  nearby dwarf spheroid at a redshift
of 0.0375$\pm$0.002.
The lensed  quasar is
at a redshift of 0.6842$\pm$0.0014 and its
multiple images are distributed almost 360$^{o}$ around
the lens nearly along a ring of radius $\sim$6."0.
Single component lens models require a mass of the galaxy of almost
10$^{12}$~M$_{\odot}$ within 6".0 from the lens center.
With the available data we are unable to determine the exact positions,
orientations and fluxes of the quasar and the galaxy, though
there appears evidence for a double or multiple merging image of the
quasar.
We have also detected strong radio
and X-ray emissions from this system. It is indicative
that this ring system may be embedded in a group or cluster of galaxies.
This unique ring, by virtue of the closeness of the
lens galaxy, offers possible probe to some of the
key issues like  mass-to-light ratio of intrinsically faint galaxies,
existence of large scale magnetic fields in elliptical galaxies etc.
\end{abstract}



\section{Introduction}
Galaxy formation  has remained an open issue in astronomy and cosmology.
Several multiwavelength deep surveys have been carried out to measure the
evolution of the mass-to-light ratio and to constrain the process(es) of
  galaxy formation and evolution, but they have their inherent limitations.
Gravitational lensing offers an unbiased technique to directly measure the
mass of galaxies as well as constraints on the mass distribution (Kochanek \& Narayan 1992; Warren et al. 1996;
Kochanek et al. 1999; Kochanek, Keeton \& McLoed, 2001; Myers et al. 2003; Warren \& Dye 2003; Bolton et al. 2004; Cabanac et al. 2007; Gavazzi et al. 2008 and references therin) 
Tight constraints on these measurements can be made
by observing nearby lens systems having multiple images as well as
perfect Einstein rings  because these systems will not
suffer from well-known ellipticity-shear degeneracy
(cf. Keeton, Kochanek, Seljak, 1997).
In addition, a detailed observation of the galaxies in the
vicinity as well as the shape of the luminous mass in the main galaxy
is feasible for very nearby galaxies.
Consequently, a strong constraint on the ratio of the mass--to--light ratio
of the lens galaxy is, in principle, possible from such a system.
For the formation of a perfect (360$^{o}$) ``Einstein Ring", the lens and
the source have to be aligned such that an extended structure
in the source straddles over at least three of the vertexes of the
tangential lens caustic. In that process the lens and an inner image
of the source will appear almost as a single object close to the center
of the ring. To date, not many Einstein rings are known that have large
circumferences (Belokurov et al. 2007 and references therein), though
recently they have discovered
  a $\sim$300$^{o}$ Einstein ring system.
Here we report the discovery of an almost perfect
optical ``Einstein Ring" (SDSS J091949.16+342304.0), a
quasar at a redshift of 0.684 lensed by a galaxy at a very low redshift
of 0.0375 forming a ring of nearly 6'' radius and associated four images.
Observations and data analysis are described in \S2. Computations of
models and interpretations of results are presented in \S3. \S4 presents
significance of this powerful lens system and discussion and conclusions are presented in \S5.

\section{Observations, data analysis  and results}
Fig. 1 show the SDSS composite image in the field of
SDSS J091949.16+342304.0.
Close-up view of the unique Einstein ring system of SDSS J091949.16+342304.0
is shown in Fig. 2. It can be seen from this figure that there appears
to be two objects at the center. This is further supported from the results
of the image-projection profile of the central region.   Nature of these
two objects has been determined from their spectra, which are shown in
Fig. 3. The SDSS spectrum of the quasar at the center
(SDSS J091949.16+342304.0) shows that it is at a redshift of 0.6842$\pm$0.0014.
This spectrum was de-reddened using the Galactic extinction curve,
(Schlegel et al. 1998), then the wavelength scale was transformed from
the observed to the source frame. This is shown in black and the SDSS
composite quasar spectrum is shown in magenta color (vanden Berk et al. 2001). We adjusted
the relative extinction between these two spectra to match their
red-wings of Mg~II (2800~\AA) line and the fluxes at 5100~\AA.
It can be seen from these two spectra that the absorption features
between 5100-5300~\AA\ region is present only in SDSS J091949.16+342304.0
and not in the composite spectrum. 
In order to check the validity of these lines, the FeII emission lines were subtracted from the observed spectrum of SDSS J091949.16+342304.0.
We obtained the  Fe~~II model spectrum in the
optical (3530 -- 7570 \AA) band  from Anabela C. Goncalves.
First, we determined the required broadening by comparing the full width at half intensity maximum (FWHM) of the iron lines in the observed and the template spectra. Then by varying the scaling factor we created a large number of template spectra, which were subtracted from the observed spectrum to determine the residual spectra. Standard deviations at the continuum around 4600~\AA\ region of the residual spectra were computed. Finally, the residual spectrum with the least value of the standard deviation was subtracted from the observed spectrum (Veron-Cetty, Veron \& Goncalves   2001; Vestergaard  \& Wilkes  2001).
The  spectrum of SDSS J091949.16+342304.0 without iron emission lines clearly displays the presence of Ca~II lines with high significances. These lines are clearly  absent in the SDSS composite  quasar spectrum, which can be seen from Fig. 3.

Next, we subtracted the SDSS composite  quasar spectrum from the observed spectrum of SDSS J091949.16+342304.0 and
the residual spectrum, at the observer's frame,  is shown at the upper plot of Fig. 4. Clearly,  two  absorption lines around 9000~\AA\ (8863 and 8988~\AA) and one around 5040~\AA\ are present. Other prominent absorption and emission fetures present in this spectrum are the artifacts.
All these three absorption lines (H$\beta$ and Ca~II)
are consistent with the redshift of 0.0375$\pm$0.002.
Thus, we identified these absorption features  as
redshifted (0.0375$\pm$0.002) H$\beta$ and Ca~II  absorption lines (8498, 8542 and 8662~\AA), which are marked on this figure. We could not identify the first Ca~II triplet line (8498~\AA), which will be redshifted at 8817~\AA\ as this line is located at the red-wing of the redshifted (0.6842$\pm$0.0014)
Mg triplet lines (Mg1, Mg2 and Mgb around 5175\AA) of the host galaxy of
the quasar. In addition, we could not identify the H$\alpha$ line from the foreground galaxy, because it is located on the blue-side  of the quasar's H$\delta$ emission line. To identify the nature of the foreground galaxy we compared the residual spectrum with the SDSS spectra of different types of galaxies, which are at redshifts between 0.037 and 0.038 and are fainter than 18.2 mag in the SDSS r-band. The limit of 18.2 mag comes from the SDSS photometric results of the central region of Fig. 2. In this figure we have marked two objects with a circle and an ellipse, whose measured brightnesses are 19.6$\pm$0.2 and 18.4$\pm$0.1 mag in the SDSS r-band, respectively. Finally, from the results of the comparison of spectra we find that the residual spectrum is similar to that of the dwarf elliptical (dE) galaxy, which is shown in the lower plot of Fig. 4. In addition, the flux and the photometric magnitude of the elliptical object of Fig. 2 is consistent with that of the dE galaxies at similar redshift. Furthermore, we found from the SDSS database that there are at least a few quasars at redshifts between 0.68 and 0.69 (SDSS J154127.26+405720.2, SDSS J155900.8+062412.0, etc.) whose radio and optical spectral properties are similar to that of SDSS J091949.16+342304.0. Photometric magnitudes of these quasars are between 19.5 and 20.0 mag. In addition, there are many quasars at redshifts between 0.68 and 0.69, whose optical spectral properties are similar to that of SDSS J091949.16+342304.0 and are fainter than 21 mag. All these photometric and spectroscopic results indicate that the central region of Figs. 1 and 2 contains a nearby dE galaxy and a quasar.
It is important to mention here that these results should be taken cautiously until future high spatial resolution, peferably with the Hubble Space Telescope, images confirm the positions, orientations and brightnesses of these objects.

In Fig. 2 we have marked two objects with ``A" and ``B". It appears that this object ``A"
could be composed of multiple images. We obtained its optical spectrum
on March 06, 2007 using the Low Resolution Spectrograph at the Nasmyth B
focus of the Telescopio Nazionale Galileo  with 2000~s
exposure (TNG is a 3.58~m optical/infrared
telescope located in the Island of San Miguel de La Palma). This observation was carried out in the Long Slit Spectroscopy
mode (LR-R Grism \#3) with a camera, which is equipped with a 2048 x 2048
Loral thinned and back-illuminated CCD. This spectrum is shown in Fig. 5 
in red color (middle plot), without corrections for the atmospheric absorptions,
which are marked with ``T" and their presence have affected the strengths
of some emission lines.  For comparison, the SDSS spectrum of the quasar
is shown in black. Five redshifted emission lines of [OII~3727~\AA], [NeIII 3869~\AA], H$\gamma$ (4340~\AA),
H$\beta$ (4860~\AA) and [OIII~5007~\AA]  that are common between these two spectra,
have been marked with  the vertical dashed lines.  
While the spectrum of object ``A" was obtained, the position of the slit of
the spectrograph was aligned in such a way that the object ``B" was on the slit.
We extracted the spectrum of object ``B" and the highly smoothed spectrum is also
shown in Fig. 5 (bottom plot in green color). Four redshifted emission lines of [OII~3727~\AA], [NeIII 3869~\AA],
[OIII~4959\AA) and [OIII~5007~\AA] are labeled in this figure
and their redshifted wavelengths are same as those of object ``A" and the SDSS quasar.
These results indicate that the central quasar and objects ``A" and ``B" are
at the same redshift (0.6842$\pm$0.0014). However, these results have to be confirmed with high signal-to-noise ratio spectra.

We also obtained the near-infrared spectra of the quasar and the object ``A"  on March 07, 2007, using the Near Infrared Camera
Spectrometer at the TNG with a HgCdTe Hawaii 1024x1024 array detector
for 500~s exposure. These spectra are shown in Fig. 6 with the upper one
being the quasar spectrum (black) and the lower one is for the object ``A"
(magenta). A broad emission feature around 1.105~$\mu$ is present in both
the spectra and we identify these features as redshifted H$\alpha$ emission line with z=0.684. 
Again, these results  suggest that the quasar and object ``A" are at the same redshift.

We searched the 2MASS database for the counterparts of the central objects of the ring system.
A central extended-blob was detected only in the J-band with  few more objects within a circle of 30" radius.
Positions of these objects coincide with the bright optical counterparts present in the SDSS images. The central objects were also
detected in the VLA/FIRST and NVSS surveys ( 1.4 GHz) with peak-flux densities at 2.15$\pm$0.13 and 2.3$\pm$0.45 mJy, respectively. We make the reasonable assumption  that the radio emission comes 
mainly from the quasar and negligible contribution from the dwarf spheroid
lens galaxy or other nearby galaxies. Then, using the VLA/FIRST flux density, the SDSS i--band magnitude and eqn. (3) of  Shen et al. (2006), we find that this quasar is a radio-loud object with radio-loudness index $\sim$1.3. Fig. 7 shows the ROSAT/PSPC image of the ring system, which has been adaptively smoothed. It can be clearly seen from this figure that bright X-ray emission is present in and around  the ring system.
This is indicative of X-ray emission from the quasar, its images and from the surroundings, which may contain X-ray emitting group of galaxies or similar objects. This image was located on the outer part of the ROSAT/PSPC detector. The PSF of the PSPC detector varies too much across the field of view and in the outer parts it 
shows strong non-symmetric features.  This did not permit us to deconvolve this image in a meaningful way (private communication with Frank Haberl). Future, high spatial resolution X-ray observations will reveal the details of this ring system. {\sl In summary, all the observed results suggest that we have detected a system, which consists of a foreground dE galaxy at a redshift of 0.0375$\pm$0.002 and, at least, three quasars (central quasar and objects ``A" and ``B") at the redshifts of 0.6842$\pm$0.0014}.

\section{Physical triple quasars or lens$?$}
Many physical binary quasars have been detected (Djorgovski et al. 1987; Meylan et al. 1990; Hennawi et al. 2006; Myers et al. 2006). However, presently, it is not clear whether these are physical binary quasars or gravitational lens systems (Kochanek et al. 1999; Mortlock et al. 1999). To date, no physical triple quasars have been unambiguously detected (Djorgovski et al. 2007; Sochting et al. 2008). All the known triple quasars associated with a galaxy or a group of galaxies or cluster of galaxies are gravitational lens systems (Kochanek \& Narayan 1992; Warren et al. 1996;
Kochanek et al. 1999; Kochanek, Keeton \& McLoed, 2001; Myers et al. 2003; Warren \& Dye 2003; Bolton et al. 2004; Cabanac et al. 2007; Gavazzi et al. 2008 and references therin)  Thus, it is strongly indicative that SDSS J091949.16+342304.0 system (Figs. 1 and 2) is a gravitational lens system. In this system we do not see the source quasar, which is most likely fainter than 21 mag. At the center we see a lensing galaxy (dE galaxy) and an image of this quasar. Objects ``A" and ``B" are the other two images of the quasar. Image ``A" is most likely composed of two or more images.
%
    \begin{figure*}
    \centering
   \includegraphics[angle=-90,scale=.50]{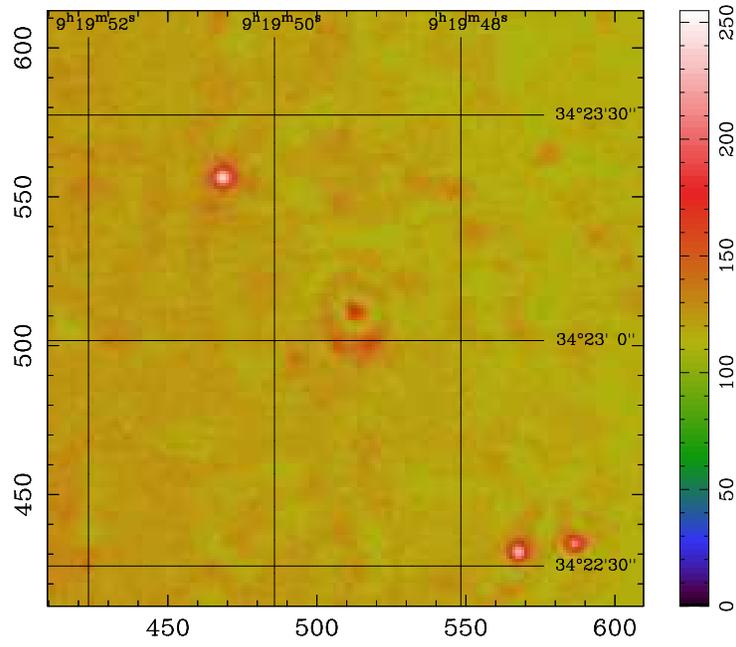}

    \caption{SDSS composite image around the Einstein ring system, SDSS J091949.16+342304.0. North is up and east to the left. The size of the image around 82".0$\times$82".0.}
               \label{Fig1}%
     \end{figure*}

%
    \begin{figure*}
    \centering
   \includegraphics[scale=.50]{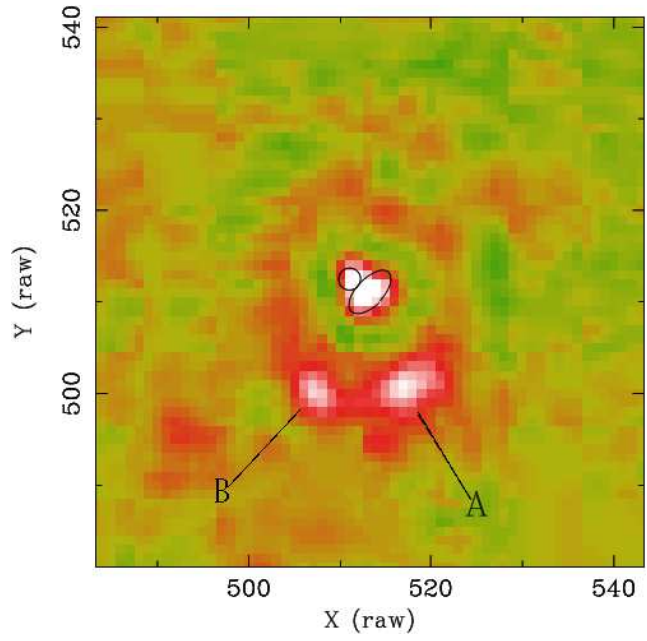}

    \caption{Same as Fig. 1, but for small size (25".0$\times$25".0). The ring is clearly visible. A small circle and the ellipse show the positions of the source and the lensing galaxy. ``A" and ``B" are the two images of the source. It appears that the image ``A" may consists of multiple images.}
               \label{Fig1}%
     \end{figure*}

%
    \begin{figure*}
    \centering
   \includegraphics[angle=-90,scale=.50]{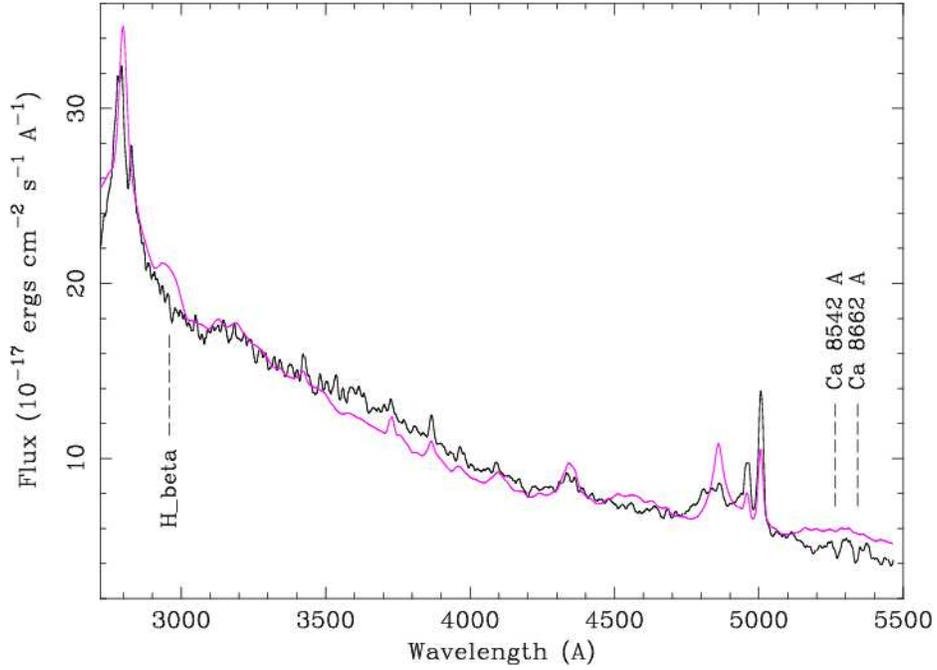}

    \caption{The Galactic extinction corrected rest-frame observed spectrum of SDSS J091949.16+342304.0, which is at the redshift of 0.6842$\pm$0.0014 (black). For comparison, SDSS composite quasar spectrum, at a redshift of 0.5, is shown in magenta color. From the comparison of the two spectra, it can be seen that the absorption lines are absent in the SDSS composite quasar spectrum. The vertical dashed lines are drawn at the position of three absorption lines, which correspond to H$\beta$ and two Ca~II lines (8542 and 8662~\AA) of the foreground galaxy, which is at a redshift of 0.0375$\pm$0.002. }
               \label{Fig2}%
     \end{figure*}

    \begin{figure*}
    \centering
   \includegraphics[angle=-90,scale=.50]{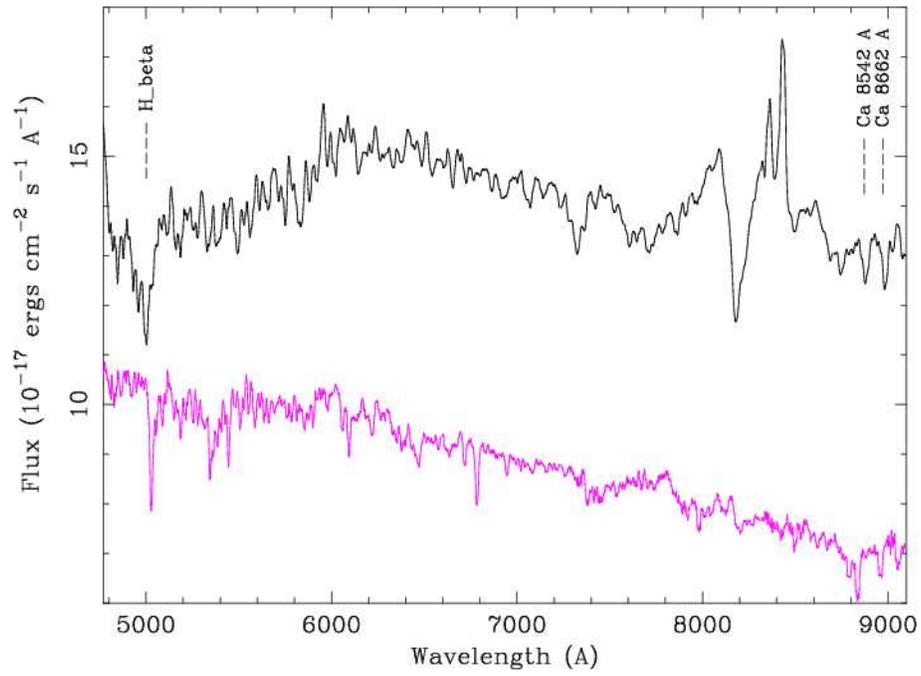}

    \caption{The upper plot shows the residual spectrum between SDSS J091949.16+342304.0 and the SDSS composite quasar spectrum. The lower plot shows the SDSS composite spectrum of dwarf elliptical galaxy at a redshift of 0.037, which we generated using the spectra from the SDSS database. H$\beta$ and two Ca~II lines (8542 and 8662~\AA) are marked with vertical dashed lines. The residual spectrum has been shifted up for clarity.}
               \label{Fig4}%
     \end{figure*}

%
    \begin{figure*}
    \centering
   \includegraphics[angle=-90,scale=.50]{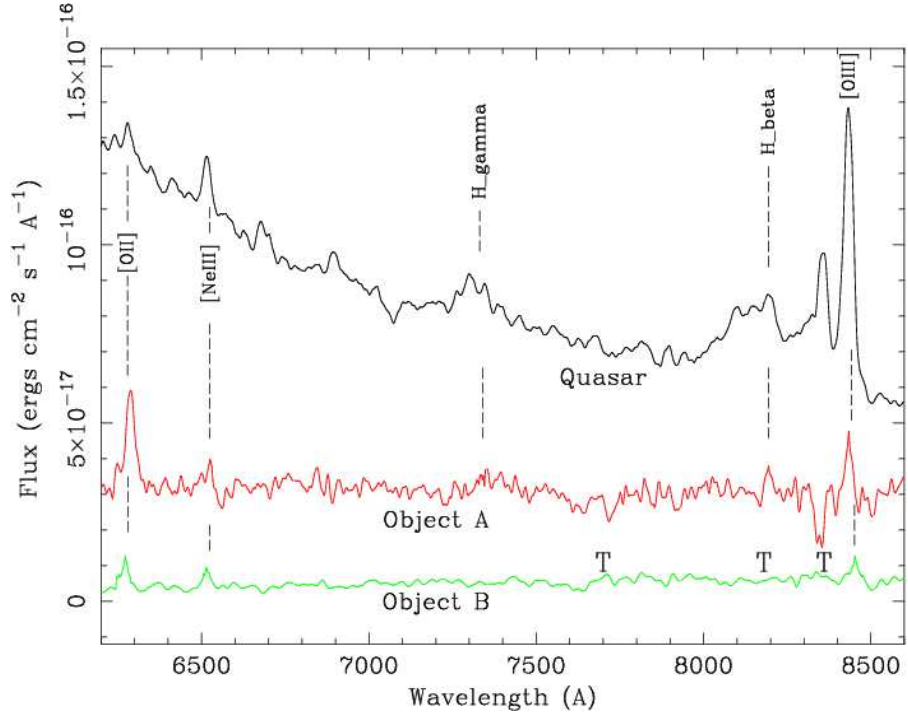}
    \caption{Observed spectra of objects ``A" and ``B", in red and green colors, respectively. at the bottom. The atmospheric telluric bands are marked with``T". For comparison, the spectrum of the quasar, SDSS J091949.16+342304.0, is shown in black at the top. Four emission lines, which are at the same positions between the three spectra are marked with vretical dashed lines. In addition, the H$\gamma$ line is marked between object ``A" and the quasar. This shows that the quasar and objects ``A" and ``B" are at the same redshift.}
               \label{Fig3}%
     \end{figure*}

%
    \begin{figure*}
    \centering
   \includegraphics[angle=-90,scale=.50]{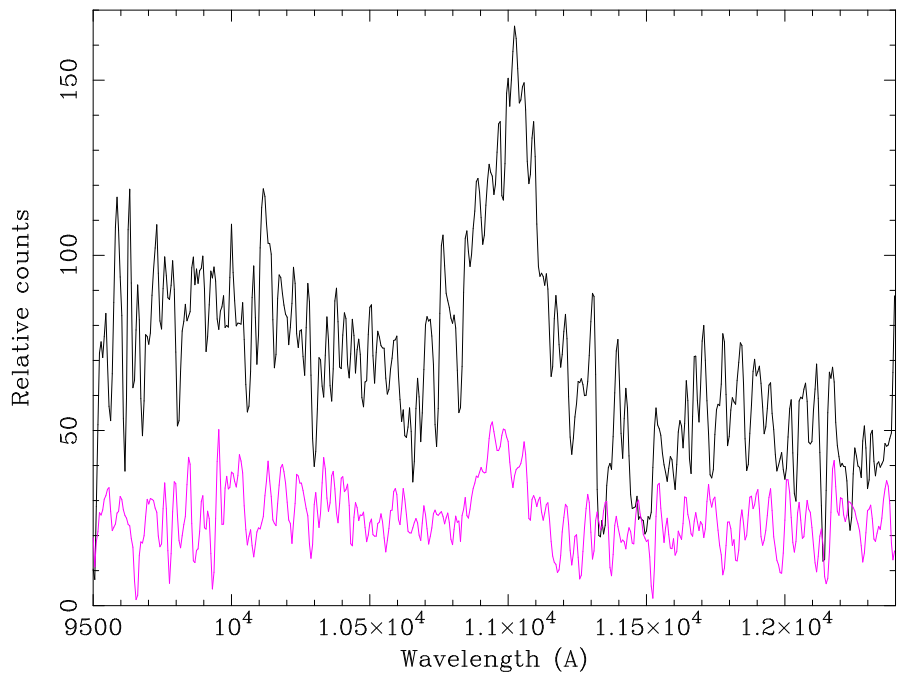}

    \caption{Observed near infrared spectra of the quasar, SDSS J091949.16+342304.0 (black), and its image, ``A" (magenta). The broad emission lines around 11000~\AA\ are the redshifted H$\alpha$ line.}
               \label{Fig3}%
     \end{figure*}

%
    \begin{figure*}
    \centering
   \includegraphics[angle=-90,scale=.50]{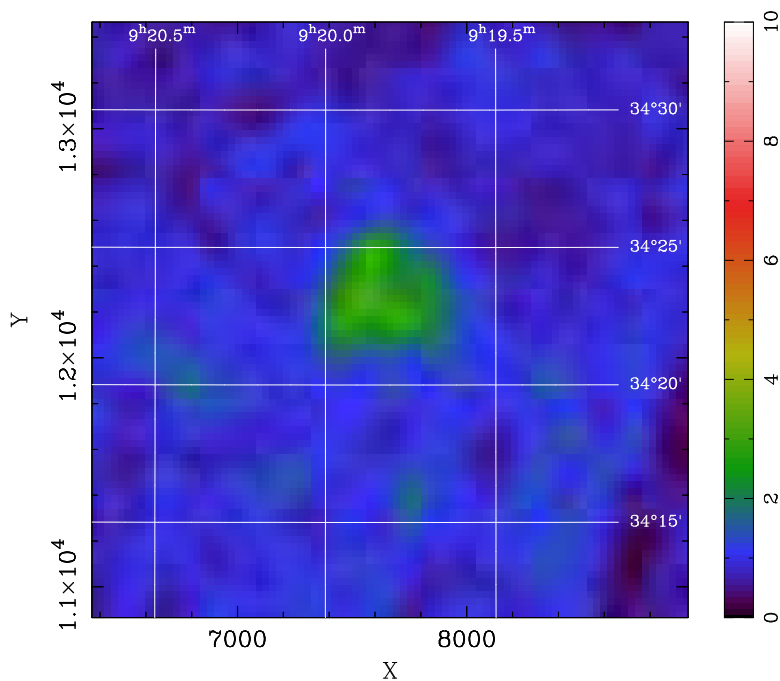}

    \caption{Adaptively smoothed ROSAT/PSPC image at the position of SDSS J091949.16+342304.0. It can be seen that the emission is extended at the arcminute scale.}
               \label{Fig4}%
     \end{figure*}

\section{Models}
The system consists of a quasar and its host at redshift of 0.6842$\pm$0.0014
lensed by a dwarf spheroid galaxy at a redshift of 0.038 forming mutiple
images of the quasar and an almost perfect 360$^{o}$ ring of
radius $\sim$6."0.
We can identify atleast three distinct images of the quasar, though their
positions are not robust.  Due to poor seeing, we cannot resolve the
images fully, but the configuration is similar to the well-studied
systems B1422+231 and Q1938+666 (cf. Narasimha \& Patnaik, 1993).
We believe that the brightest image, ``A", is probably a double,
which should be resolvable
under improved signal and a seeing of better than 1 arcsec.
The position of an image at arcsecond separation from the centre of
the lens galaxy is still very uncertain and  its flux
cannot be determined accurately with the present set of data
since the galaxy at a redshift of 0.038 is much brighter than
any of the images. 

The lens is at a very low redshift and hence its surface mass density is
not much greater than the critical value for multiple imaging.
Consequently, the central region of the lens galaxy turns out to be important
in determining the characteristics of the inner image.
The profile of the Einstein ring of radius 6".0 is determined by the
large scale shear produced
by the main galaxy or any galaxy groups. Consequently, a good map of the
image configuration could render this system a valuable probe of the
dynamical mass distribution in elliptical galaxies and at larger scales.

\subsection{Model1:Model independent minimal limts for lens mass:}

  A single component spherical mass
  at a redshift of 0.0375 (distance of the order of 160 Mpc)
acting as gravitational lens for a background source of redshift 0.68
and producing three images at approximately 6 arcsecond from the lens
centre should have a minimum mass given by

  $$M_{min} = {{D_{eff} c^2 \theta_E^2}\over {4 G}}$$
where $D_{eff}$ = ${{D_S D_L}\over D_{LS}}$, the combination of distances to
  the source D$_S$, to the lens D$_L$ and the distance from the source
to the lens D$_{LS}$ is essentially the distance to the lens in this case.
c=velocity of light, G=Gravitational
constant and $\theta_E$ is the Einstein radius, which is about 5 arcseconds.
Consequently, the minimum mass in the absence of large scale shear
due to the outer regions of the galaxy or additional galaxies
  is of the order of 9$\times$ 10$^{11}$ M$_\odot$.
However, the ellipticity of the
lens can reduce the minimum surface mass required for multiple image formation
(Subramanian \& Cowling, 1986). 
At present  we cannot determine the ellipticity of the lens or importance
of external shear, without an observational information about the
orientationalong which the double images in A merge or their substructures.
Consequently, we have constructed  a model for a spheroidal lens with
ellipticity to match the approximate position of the images ``A" and ``B"
and the apparent direction along which two subimages appear to be merging
in image ``A" 
(Narasimha, Subramanian \& Chitre, 1982).
In Figure 8, a single lens produces the essentials of the model, though
due to limitations of data the model is just indicative. For instance,
the high eccentricity of 0.9 used
to get the curvature of the ring near the images will change
with better quality data on the images and then, the mass of the lens could
increase.
In this model, the central cusp like mass profile is simulated through
a 200 pc bulge of mass 10$^{10}$ solar mass and a truncated King profile
has core radius of 1.2 kpc, eccentricity of 0.9 and mass
of 6$\times$ 10$^{11}$ M$\odot$ as expected from the simplistic considerations.
For the observed luminosity of the order of 10$^9$ L$\odot$,
the Mass to Light ratio is substantial.
Such a model cannot be completely ruled out, specially in view of the
massive low surface brightness galaxies observed (Impey et al, 1988;
Bothun et al, 1997),
but from the observed optical flux distribution,
 the lens galaxy is unlikely to be of that type.

%
    \begin{figure*}
    \centering
    \includegraphics[angle=0,scale=.70]{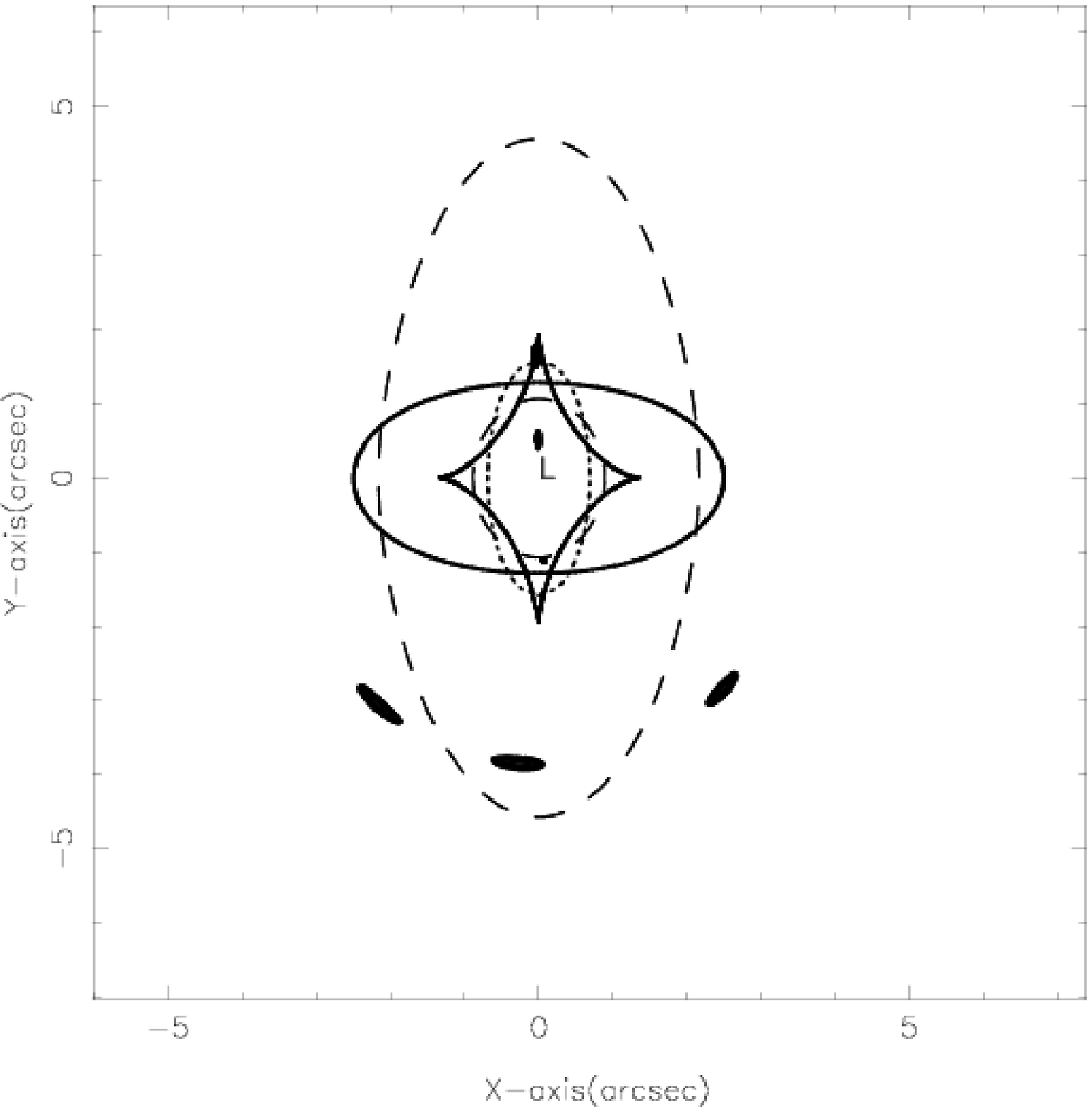}
    \caption{Image configuration due to the lensing action of a spheroidal 
mass distribution. The solid lines are the critical curves in the source
plane delineating regions of image multiplicities while the long dashed
lines are the corresponding caustics in the image plane. The 
tangnetial magnification of images tend to infinity for a pair of images
when the source lies on the diamond 
shaped critical curve and they will appear to be merging.
  The shape and position of this critical line is dependent on
the large scale shear of the lens. The outer oval shaped long dashed curve 
is its counterpart in the image plane.
The radial magnification tends to infinity for image pairs when
the source lies on the oval shaped solid line and the radially merging
images lie on the inner dashed caustic.
The dotted oval shaped curve is the spheroidal lens. The source
(a small circle just inside the diamond shaped critical curve)
has an extended arc like image consisting of two merging images,
additional external image as well as an inner image apprx. 1 arcsecond
from the lens centre. A fifth very dim image is not shown.}
       \label{nasalo2}%
     \end{figure*}

%


However,
a configuration with three images forming almost an extended arc
and another image close to the centre of the main lensing galaxy can be
a natural result when shear due to many galaxies en route to a background
source qualitatively modify the lensing action due to a foreground galaxy.
Formation of multiple images and arc--like features 
due to shear dominated
gravitational lens action is discussed by Narasimha (1993) and
Narasimha \& Chitre (1993), who studied the effects of the
central mass and the large scale shear when the lens can barely form
multiple images.
The essential features of such a configuration can be studied by taking
the  gravitational lens action of a spherical mass along with a constant
shear. This will reduce the mass of the main lensing galaxy,
and the ellipticity of the model is a natural consequence of
the shear due to surrounding mass distribution.
But we cannot get any idea of the shear upto when we isolate the
central faint image and resolve the double image A.
However, we should add that such configurations are 
likely to be
detectable in surveys extending to fainter levels because
at arcsecond scales the gravitational pull due to multiple galaxies
at cosmological distances en route to a distant source become
important.


In our system, there is indeed evidence for the existence of
at least three groups of galaxies along three directions
around the observed foreground galaxy. We believe that their combined
action is likely to be responsible for the observed multiple images.
When two lenses of equal strength are located at two vertices of
a triangle and additional lenses are distributed along the third direction,
the midpoint of the first two lenses and surrounding regions have
nearly constant shear due to the comined action of the lenses.
This scenario is not as unlikely as it might appear:
The central region of a weak Group of galaxies could produce this
configuration or meeting points of filaments in large scale structures
have similar morphology. Even by pure conincidence, sometimes large number of
galaxies can simulate this effect since the bending angle drops very slowly
as the inverse of distance while the number of lenses that can act
at a point, which is proportional to the area of the region, increases
as the square of the distance to the line of sight to the background source.
{\it If indeed this is the scanario, it will affect both inferences of
cosmological matter power density based on multiply--imaged large
separation lenses as well as estimation of masses of the individual lenses}.

\section{Significance of SDSS J091949.16+342304.0 as a diagnostic tool to probe
mass inhomogeneities at small and large scales:}

We have only preliminary data for this powerful lens system and 
consequently, the model given in the previous section should be treated 
as illustrative. But the shear dominated lens system
and its diagnostic power should not be overlooked, while analysing
images from deep surveys.
We could speculate some of the results expected from a detailed 
multifrequency study of this system.

\subsection{ Large scale mass distribution in the main lens:}

  The object appearing as the main lens is a dwarf spheroid of R magnitude 18
  at a distace of apprx. 160 Mpc and hence abs R mag of -18.1.
  It would be very difficult to have reliable direct observation and analysis
  of such an object in a lens system if it were, say, at a redshift of 0.3
  because the lens will be typically 5 mag fainter and the images
  of the background source will dominate at almost al wavelengths.
  However, for the present system we have the lens five times brighter than 
the background images in optical bands and extends over a few arcseconds.
  At a distance of 160 Mpc, 1 arcsecond corresponds to about 800 pc
  and hence, with deeper optical and infrared images, we can obtain high
dynamical range images of the galaxy to determine the scale length at 
which the flux from the lens decreases. A good model of the lens system 
can independetly give an idea of how the gravitational mass drops off,
  if we have good radio images of the system showing the details of the
  curvature of the Einstein Ring at subarcsecond scales or we can
map the images at radio to determine the shape 
of the merging images.

At this distance, galaxy clusters of even moderate masses can be
studied in X-ray and the temperature of the intracluster gas can be
estimated.
  If we have good X--ray map showing the massive objects in the field and their
  temperature, we can determine the mass distribution at
100s of kpc corresponding to arcminute scale or rule out the 
possibility of any cluster of galaxies associated with the main lens.
If indeed we are able to detect {\it any warm X--ray corona at a few tens 
of 
arcseconds which can satisfactorily explain the observed image 
configuration, 
it might open up the possibility of dwarf--like galaxies being centres
of massive dark elliptical halos.}

\subsection{Small scale mass distribution:}

At present we have limited information about the inneer image, though we
can determine its separation from the lens centre of 0.85 arcsecond.
However, further details should await better quality data.
If we have accurate position of this image and magnification with respect 
to image ``B", it could provide some constraints on the cusp like mass 
distribution at the centre of the dwarf galaxy as well as possible 
presence of a bulge component at a kiloparsec scale.

Ideally, we expect image ``A" to be a double and from an inspection of the 
available images we feel they are separated by 0."5-0."7. 
  If radio, optical and X-ray observations  confirm this hypothesis
  and their relative flux ratios are consistent, we should expect a smooth
  mass distribution at scales of 1 kpc in the lens.
  But a million solar mass globular cluster at 160 Mpc has an Einstein radius
  of 6 milliarcsec. Consequently, any possible mass inhomogeneity
  of this scale, along with the main lens can introduce microimages
  separated by a few tens of milliarcsecond and their signature can be
  seen in the high-reoslution radio image. Certainly, a high sensitivity radio image of the
  Einstein Ring should show the signatures of such mass inhomogeneity,
  mainly due to the proximity of the lens.
  We feel that this system, due to its proximity, radio loudness and the
  Einstein Ring of large radius is an ideal candidate to probe inhomogeneities
  at tens of millions of solar mass at 100 parsec scale.

\subsection{Role of other massive galaxies in the field:}

It is not clear if many other galaxies within an arcminute are part a
Group or Cluster of galaxies or other normal field galaxies at various 
redshifts which happen to be near the line of sight to the Einstein Ring.
The radius of the Einstein Ring is not small, and we do expect
a few tens of galaxies in the field within 10$^3$ arcsec$^2$ area.
Consequently, statistics does not help us differentiate between the 
possibilities in the absence of redshift or X-ray flux measurements.
But, even if the galaxies are chance coincidence, their role
in the formation of the Einstein Ring is important.
   The bending angle drops off as 1/b where b is the impact parameter
of the photon path with respect to any such galaxy.
  The number of lenses in the plane of the sky will increase as the square
  of the impact parameter, if the lenses are homogeneously distributed.
  Consequently, if there is an extra concentration of mass along some
  direction even at tens of arcsecond away, it can have noticeable
  efect on the image formation, which might dominate over the gravitational
  effects of the main dwarf galaxy lens.
This might not have been the case for a typical  giant
  elliptical galaxy at a redshift of 0.3  acting as the lens, producing
an arcsecond scale ring mainly due to the central mass of the galaxy
and shear due to its large scale mass. For the present lens configuration,
even if there is no extra concentration along a specific direction,
  we can still have a fairly wide region where an almost constant shear due to
  many of these galaxies act. This appears to be likely
  because there are about a dozen or more galaxies at ten to 30 arcsecond 
away.
  We do not have redshift or other details of these gaalxies,
  but from an inspection of the position and brightness of these galaxies,
the Einstein Ring appears to be almost at the centroid of a triangle 
formed by these galaxies. If indeed this is {\it a chance coincidence and 
we notice it only due to the proximity of the main lens, the possibility 
of many of the multiple image sysetms reported in the literature
  being artefact of the
  specific large scale galaxy distribution along their line of sight}
  should be taken into account while constructing their model as well as
  estimating the amplitude of matter power fluctuation based on
  the statistic of image sepration in strong lens systems.

\section{Discussion and conclusions}

We have discovered a gravitational lens system consisting of possible four
images of a quasar and an almost perfect Einstein Ring of radius nearly 6''.
  The quasar has a redshift of 0.684 and the main lens appears to be
a 18.1 magnitude galaxy at a redshift of 0.0375.
Since the lens is at a very low redshift and the source  is
radio-loud  and X--ray luminous, the system provides a powerful tool to study
mass distribution within lens galaxy, specially in the central regions.
A detailed observation and analysis of the system could provide many direct
tests or cinfirmations in lensing as well as strcuture of galaxies:
e.g. the ring morphology as well as the imaging of the field of galaxies around
could give an indication of the importance of Groups of Galaxies at even very
low redshifts as powerful lenses, possible chance conincidences
producing many of the eye--catching lens configurations, mass to light ratio
of lens galaxy as a function of radial distance,
possible existence of large scale magetic fields in the elliptical lens galaxy
are some of the important problems that can be addressed with this system
simply because of its proximity, scale and being loud in radio, optical
as well as X--ray.

Though we have only preliminary data for this system, it could have
some important implications to cosmolgy:

    \begin{enumerate}
       \item If  a single galaxy is the main lens, it should have a
mass to light ratio of upwards of 500 M$_{\odot}$/L$_{\odot}$.
This will possibly be a new result for a dwarf spheroid of similar
luminosity.

       \item If the lens consists of a Group of galaxies, the possibility of
some of the weak groups of galaxies having high surface mass density
to produce multiple images and extended arcs even at a very low redshift
of 0.0375 should be considered while estimating the matter power density
from a surveylike SDSS, even though those groups may not be conspicous
in optical luminosity. In this context, the lensing due to
filaments along favourable directions should be taken into account
while estimating cosmological parameters, for instance, from
cosmic shear data.

       \item If the lens shear is due to chance location of many galaxies
along the line of sight to a distant background source, this fact should be
taken into account while using gravitational lens systems of large angular
separation to estimate the mass of very massive objects. This will have
far reaching implications when gravitational lens is used to calibrate,
for instance, masses of galaxy--clusters and hence, the $\sigma_8$
parameter for the amplitude of matter power is derived.

    \end{enumerate}

\acknowledgments
Our sincere thanks to the referee for valuable comments and suggestions that helped to improve the paper.
We thank Carlos M Gutierrez de la Cruz and Martin Lopez-Corredoira who obtained the optical and near-infrared spectra of the quasar and its images, presented in this paper, during their observations.
In this paper, we have extensively used data from the Sloan Digital Sky Survey (SDSS). Funding for the SDSS has been provided by the Alfred P. Sloan Foundation, the
Participating Institutions, the National Aeronautics and
Space Administration, the National Science Foundation,
the U. S. Department of Energy, the Japanese Monbuka-
gakusho, and the Max Planck Society. The SDSS website is
http://www.sdss.org/.
  This research has made use of the NASA/IPAC Extragalactic Database
(NED) which is operated by the Jet Propulsion Laboratory, California Institute of Technology, under contract with NASA; of data products from the Two Micron All Sky Survey, VLA/FIRST, NVSS and ROSAT/PSPC.

\end{document}